\documentclass{giq7}
\usepackage{amsmath,amssymb}
\def\ri{{\mathrm i}}
\def\rd{{\mathrm d}}
\def\re{{\mathrm e}}
\def\a{{\boldsymbol a}}

\def\c{{\boldsymbol c}}
\def\d{{\boldsymbol d}}

\def\ad{\mbox{ad\,}}
\def\Ad{\mbox{Ad\,}}
\def\tr{\mbox{tr\,}}

\def\otimescomma{\mathop{\otimes}\limits_{'}}
\def\wedgecomma{\mathop{\wedge}\limits_{'}}

\def\bbbe{{\Bbb E}}

\def\bbbz{{\Bbb Z}}
\def\openone{\leavevmode\hbox{\small1\kern-3.3pt\normalsize1}}

\begin{document}

\title[NEW MULTI-COMPONENT NLS TYPE EQUATIONS ON SYMMETRIC SPACES]
{NEW INTEGRABLE MULTI-COMPONENT NLS TYPE EQUATIONS ON SYMMETRIC SPACES:\\
$\bbbz_4 $ AND $\bbbz_6 $ REDUCTIONS}
\author[G. G. Grahovski, V. S. Gerdjikov,
N. A. Kostov \MakeLowercase{and} V. A. Atanasov]{Georgi G. Grahovski$^{\dag}$, Vladimir S.
Gerdjikov$^{\dag}$, Nikolay A. Kostov$^{\ddag}$
\MakeLowercase{and} Victor A. Atanasov$^{\dag}$ }
\address{$^{\dag}$ Institute for Nuclear Research and Nuclear Energy,
Bulgarian Academy of Sciences, 72 Tsarigradsko chaussee, 1784
Sofia, Bulgaria \\
$^{\ddag}$ Institute of Electronics, Bulgarian Academy of
Sciences, 72 Tsarigradsko chaussee, 1784 Sofia, Bulgaria}

\begin{abstract}
The reductions of the multi-component nonlinear Schr\"{o}dinger \-
(MNLS) type models related to {\bf C.I} and {\bf D.III} type
symmetric spaces are studied. We pay special attention to the MNLS
related to the $\mathfrak{sp}(4) $, $\mathfrak{so}(10)$ and
$\mathfrak{so}(12)$ Lie algebras. The MNLS related to $
\mathfrak{sp}(4)$ is a three-component MNLS which finds applications to
Bose-Einstein condensates. The MNLS related to $\mathfrak{so}(12)$  and
$\mathfrak{so}(10) $ Lie algebras after convenient $\bbbz_6 $ or $\bbbz_4
$ reductions reduce to three and four-component MNLS showing new types of
$\chi ^{(3)} $-interactions that are integrable. We briefly explain how
these new types of MNLS can be integrated by the inverse scattering
method. The spectral properties of the Lax operators $L $ and the
corresponding recursion operator $\Lambda $ are outlined. Applications to
spinor model of Bose-Einstein condensates (BEC's) are discussed.
\end{abstract}

\maketitle

\section{Introduction}\label{sec:1}

When spinor BEC's are trapped in magnetic potential, the spin
degree of freedom is frozen. However, in the condensate trapped by
an optical potential, the spin is free. We consider BEC's of
alcali atoms in the $F=1$ hyperfine state, elongated in $x$
direction and confined in the transverse directions $y,z$ by
purely optical means. Then, in the absence of external magnetic
fields is characterized by the magnetic quantum number which has
three allowed values $m_{F}=1,0,-1$.Thus the assembly of atoms in
the $F=1$ hyperfine state can be described by a normalized spinor
wave function ${\bf\Phi}(x,t)=(\Phi_1(x,t),\Phi_0(x,t),
\Phi_{-1}(x,t))^{T}$ whose components are labelled by the values
of $m_F $.  In short the dynamics of such BEC's is described by a
three-component Gross-Pitaevskii (GP) system of equations. In the
one-dimensional approximation described above the GP system goes
into the following  three-component nonlinear Schr\"{o}dinger
(MNLS) equation in (1D) $x$-space \cite{IMW04}:
\begin{eqnarray}\label{eq:1}
&& \i\partial_{t} \Phi_{1}+\partial^{2}_{x}
\Phi_{1}+2(|\Phi_{1}|^2
+2|\Phi_{0}|^2) \Phi_{1} +2\Phi_{-1}^{*}\Phi_{0}^2=0, \nonumber \\
&& \i\partial_{t} \Phi_{0}+\partial^{2}_{x}
\Phi_{0}+2(|\Phi_{-1}|^2
+|\Phi_{0}|^2+|\Phi_{1}|^2) \Phi_{0} +2\Phi_{0}^{*}\Phi_{1}\Phi_{-1}=0,\\
&& \i\partial_{t}\Phi_{-1}+\partial^{2}_{x}
\Phi_{-1}+2(|\Phi_{-1}|^2+ 2|\Phi_{0}|^2) \Phi_{-1}
+2\Phi_{1}^{*}\Phi_{0}^2=0. \nonumber
\end{eqnarray}
This model is integrable by means of inverse scattering transform
method \cite{IMW04}. It also allows an exact description of the
dynamics and interaction of bright solitons with spin degrees of
freedom. Matter-wave solitons are expected to be useful in atom
laser, atom interferometry and coherent atom transport. It could
contribute to the realization of quantum information processing or
computation, as a part of new field of atom optics. Lax pairs and
geometric interpretation of the model (\ref{eq:1}) are given in
\cite{ForKu*83}. Darboux transformation for this special
integrable model is developed in \cite{LLMML05}. We will show that
the system (\ref{eq:1}) is related to symmetric space ${\bf
C.I}\simeq {\rm Sp(4)}/{\rm U(2)}$ with canonical
$\bbbz_2$-reduction and has natural Lie algebraic interpretation.

The applications of the differential geometric and Lie algebraic
methods to soliton type equations lead to the discovery of close
relationship between the MNLS equations and the symmetric spaces
\cite{ForKu*83}. It was shown that these MNLS systems have Lax
representation with  the generalized Zakharov--Shabat system as
the Lax operator:
\begin{eqnarray} \label{eq:Lax-MNLS}
L\psi (x,t,\lambda ) \equiv \i {\rd \psi \over \rd x} + (Q(x,t) -
\lambda J)\psi (x,t,\lambda )=0,
\end{eqnarray}
where $J $ is a constant element of the Cartan subalgebra
$\mathfrak{h} \subset \mathfrak{g} $ of the simple Lie algebra
$\mathfrak{g} $ and $Q(x,t) \equiv [J,\widetilde{Q}(x,t)]$. In
other words $Q(x,t) $ belongs to the co-adjoint orbit
$\mathcal{M}_J $ of $\mathfrak{g} $ passing through $J$.

The choice of $J $ determines the dimension of $\mathcal{M}_J $
which can be viewed as the phase space of the relevant nonlinear
evolution equations (NLEE). It is equal to the number of roots of
$ \mathfrak{g} $ such that $\alpha (J)\neq 0 $. Taking into
account that if $\alpha $ is a root, then $-\alpha $ is also a
root of $\mathfrak{g} $; therefore $\dim \mathcal{M}_J $ is always
even.

We concentrate on those most degenerate choices for $J $ for which
$\ad_J $ has just two non-vanishing eigenvalues $\pm 2a $; in this
case $J^2=a^2\openone$. Such choices of $J $ are compatible with
several types of symmetric spaces: ${\bf A.III} \simeq {\rm
SU(p+q)}/{\rm S(U(p)}\otimes {\rm U(q)})$, ${\bf C.I}\simeq {\rm
Sp(2p)}/{\rm U(p)}$ and ${\bf D.III}\simeq {\rm SO(2p)}/{\rm
U(p)}$ \cite{ForKu*83,Helg}. The classification of the symmetric
spaces related to a given simple Lie algebra $\mathfrak{g}$ is
directly related to the classification of the Cartan involutions
(i.e. to the classifications of the real forms) that the algebra
admits. For more details see e.g. \cite{Helg,Loos}.

The interpretation of the ISM as a generalized Fourier transforms
and the expansion over the so-called ``squared'' solutions (see
\cite{K,G*86} for regular and \cite{VSG*94,MV,VSG-pres} for
non-regular $J$) allow one to study all the fundamental properties
of the corresponding NLEE's. These include: i) the description of
the class of NLEE related to a given Lax operator $L(\lambda)$ and
solvable by the ISM; ii) derivation of the infinite family of
integrals of motion; and iii) their hierarchy of Hamiltonian
structures.

The degeneracy of $J $ means that the subalgebra $\mathfrak{g}_J
\subset \mathfrak{g} $ of elements commuting with $J $ (i.e., the
kernel of the operator $\ad_J $) is non-commutative which makes
more difficult the derivation of the fundamental analytic
solutions (FAS) of the Lax operator (\ref{eq:Lax-MNLS}) and the
construction of the corresponding (generating) recursion operator
$\Lambda$. Here we continue our studies in \cite{G*86,VSG*94}
finding new algebraic reductions of MNLS equations related to {\bf
C.I} and {\bf D.III} type symmetric spaces. Some of them like eq.
(\ref{eq:1}) find applications to Bose-Einstein condensates and
nonlinear optics. The derived reduced MNLS system seem to be new
to the best of our knowledge.

The present article is organized as follows: In section 2 we give
some preliminaries about the simple Lie algebras and the general
form of the MNLS models and the relevant recursion operators.
Section 3 is devoted to the spectral properties of the Lax
operator $L $. In section 4 we discuss the Hamiltonian properties
of the MNLS systems using the classical $R $-matrix method. In
section 5 we apply the approach in \cite{Za*Mi,MV} and derive the
dressing factors and the soliton solutions for symmetric spaces
related to ${\bf D}_5$ and ${\bf D}_6$ Lie algebras. Section 6 is
devoted to the analysis of the reductions of the MNLS equations by
applying the reduction group \cite{2} method. The relation between
reductions and scattering data for $L$ operator are outlined in
section 7.

\section{Preliminaries. }\label{ssec:2.1}

\subsection{Simple Lie algebras.}\label{ssec:}

Here we fix the notations and the normalization conditions for the
Cartan-Weyl generators $\{h_k, E_\alpha \}$ of $\mathfrak{g} $
($r=\mbox{rank}\,\mathfrak{g}$) with root system $\Delta$. We
introduce $h_k\in \mathfrak{h} $, $k=1,\dots,r $ as the Cartan
elements dual to the orthonormal basis $\{e_k\}$ in the root space
$\bbbe^r $ and the Weyl generators $E_\alpha $, $\alpha \in \Delta
$. Their
commutation relations 
\begin{eqnarray}\label{eq:31.2}
&& [h_k,E_\alpha ] = (\alpha ,e_k) E_\alpha , \quad [E_\alpha
,E_{-\alpha }]={2 \over (\alpha ,\alpha ) } \sum_{k=1}^{r} (\alpha
,e_k) h_k , \nonumber\\
&& [E_\alpha ,E_\beta ] = \left\{ \begin{array}{ll} N_{\alpha
,\beta } E_{\alpha +\beta } \quad & \mbox{for}\; \alpha +\beta \in
\Delta \\ 0 & \mbox{for}\; \alpha +\beta \not\in \Delta \cup\{0\}.
\end{array} \right.
\end{eqnarray}
Here $\vec{a}=\sum_{k=1}^{r}a_k e_k $ is a $r $-dimensional
vector dual to $J\in \mathfrak{h} $ and $(\cdot ,\cdot) $ is the
scalar product in $\bbbe^r $. The normalization of the basis is
determined by:
\begin{eqnarray}\label{eq:32.1}
&& E_{-\alpha } =E_\alpha ^T, \qquad \langle E_{-\alpha
},E_\alpha\rangle
={2 \over (\alpha ,\alpha ) }, \qquad 
 N_{-\alpha ,-\beta } = -N_{\alpha ,\beta }, 
,
\end{eqnarray}
where $N_{\alpha,\beta }= \pm (p+1) $ and the integer $p\geq 0 $
is such that $\alpha +s\beta \in\Delta $ for all $s=1,\dots,p $, $
\alpha +(p+1)\beta \not\in\Delta $ and $\langle \cdot, \cdot
\rangle $ is the Killing form of $\mathfrak{g}$, see \cite{Helg}.
The root system $\Delta $ of $\mathfrak{g} $ is invariant with
respect to the group $W_{\mathfrak{g}} $ of Weyl reflections
$S_\alpha $,
\begin{equation}\label{eq:32.2}
S_\alpha \vec{y} = \vec{y} - {2(\alpha ,\vec{y}) \over (\alpha
,\alpha )} \alpha , \quad \alpha \in \Delta .
\end{equation}
With each reflection $S_\alpha $ one can relate an internal
automorphism of the algebra $ \Ad_{A_\alpha}\in
\mbox{Aut}_0\,\mathfrak{g} $ which act in a natural way on the
Cartan-Weyl basis, namely:
\begin{eqnarray}\label{eq:32.3}
&& S_\alpha (H_\beta ) \equiv A_\alpha H_\beta A^{-1}_{\alpha } =
H_{\beta '}, \qquad \beta '=S_\alpha \beta, \nonumber\\
&& S_\alpha (E_\beta ) \equiv A_\alpha E_\beta A^{-1}_{\alpha } =
n_{\alpha ,\beta } E_{\beta' }, \qquad n_{\alpha ,\beta }=\pm 1.
\end{eqnarray}
Since $S_\alpha ^2=\openone $ we must have $A_\alpha ^2=\pm
\openone $.

As we already mentioned in the Introduction the MNLS equations
correspond to Lax operator (\ref{eq:Lax-MNLS}) with non-regular
(constant) Cartan elements $J\in \mathfrak{h}$. If $J$ is a
regular element of the Cartan subalgebra of $\mathfrak{g}$ then
$\ad_J$ has as many different eigenvalues as is the number of the
roots of the algebra and they are given by $a_j=\alpha_j(J)$,
$\alpha_j\in \Delta$. Such $J $'s can be used to introduce
ordering in the root system by assuming that $\alpha >0 $ if
$\alpha (J)>0 $. In what follows we will assume that all roots for
which $\alpha (J)>0 $ are positive.

Obviously we can consider the eigensubspaces of $\ad_J $ as
grading of the algebra $\mathfrak{g} $. In what follows we will
consider symmetric spaces related to maximally degenerated $J $,
i.e. $\ad_J $ has only two non-vanishing eigenvalues $ \pm 2a $.
Then $\mathfrak{g}$ is split into a direct sum of the subalgebra
$\mathfrak{g}_0 $ and the linear subspaces $\mathfrak{g}_{\pm } $:
\begin{eqnarray}\label{eq:KF14.1}
\mathfrak{ g} &=& \mathfrak{ g}_0 \oplus \mathfrak{ g}_+ \oplus
\mathfrak{g}_{-},\qquad \mathfrak{ g}_{\pm } = \mbox{l.c.}
\left\{ X_{\pm j}\,|\, [J, X_{\pm }]=\pm 2a X_{\pm } \right\}
\nonumber
\end{eqnarray}
The subalgebra $\mathfrak{ g}_0 $ contains the Cartan subalgebra
$\mathfrak{h} $ and also all root vectors $E_{\pm \alpha }\in
\mathfrak{g} $ corresponding to the roots $\alpha $ such that
$(\vec{a},\alpha )=0 $. The root system $\Delta $ is split into
subsets of roots $\Delta = \theta _0\cup \theta _+\cup (-\theta
_+)$, where:
\begin{eqnarray}\label{eq:KF14.3}
\theta _0 &=& \left\{ \alpha \in \Delta \,|\, \alpha (J) =0
\right\}, \quad \theta _+ = \left\{ \alpha \in \Delta \,|\, \alpha
(J) =a >0 \right\}.
\end{eqnarray}
We can use the gauge transformation commuting with $J $ to
simplify $Q $; in particular we can remove all components of $Q $
in $\mathfrak{ g}_0 $; effectively this means that our $Q(x,t)=
Q_+(x,t) + Q_{-}(x,t)\in \mathfrak{g}_+\cup \mathfrak{g}_- $ can
be viewed as a local coordinate in the co-adjoint orbit
$\mathcal{M}_J\simeq \mathfrak{g}\backslash \mathfrak{g}_0 $:
\begin{eqnarray}\label{eq:KF14.4}
Q_+(x,t) = \sum_{\alpha \in\theta _+}^{} q_\alpha(x,t) E_{\alpha
}, \qquad Q_{-}(x,t) =\sum_{\alpha \in \theta _-}^{} p_\alpha(x,t)
E_{\alpha }.
\end{eqnarray}
Obviously $Q_\pm \in \mathfrak{g}_\pm $ and:
\begin{equation}\label{eq:KF14.6}
\ad_J Q \equiv [J,Q] = 2a (Q_+ - Q_{-}), \qquad (\ad_J)^{-1} Q =
 {1 \over 2a} (Q_+ - Q_{-});
\end{equation}
besides $[E_\alpha ,E_\beta ]=0 $ for any pair of roots $\alpha
,\beta \in \theta _+$. This simplifies solving the recursion
relations and the explicit calculation of the recursion operator
$\Lambda $.

\subsection{Lax representation of the MNLS type models
}\label{ssec:2.2}

The operator (\ref{eq:Lax-MNLS}) together with the corresponding
operator $M(\lambda)$:
\begin{equation}\label{eq:Mop}
 M(\lambda)\psi \equiv \left(\i{\d\over \d t} - [Q, \ad_J^{-1}Q] + 2\ri
 \mbox{ad}_J^{-1}Q_x +2\lambda Q - 2\lambda ^2 J \right)
 \psi (x,t,\lambda )=0,
\end{equation}
where $Q=Q(x,t)$, provide the Lax representation for the MNLS type
systems. The compatibility condition $[L(\lambda),M(\lambda)]=0$
of (\ref{eq:Lax-MNLS}) and (\ref{eq:Mop}) gives the general form
of the MNLS equations on symmetric spaces:
\begin{equation}\label{eq:MNLSeq}
{\i\over 2} \left[J,{\partial Q \over \partial t }\right] +
{\partial^2 Q \over \partial x^2 } -2a^2
[\ad_J^{-1}Q,[\ad_J^{-1}Q,Q]]=0.
\end{equation}
Following \cite{AKNS} one can consider more general $M $-operators
of the form:
\begin{equation}\label{eq:M-op}
M(\lambda)\Psi\equiv \i {\rd \Psi \over \rd t}+\left(
\sum_{k=1}^{N}V_k(x,t)\lambda^k \right)\Psi(x,t,\lambda)=0, \qquad
f(\lambda ) = \lim_{x\to\pm\infty } V(x,t,\lambda ).
\end{equation}
The Lax representation $[L(\lambda), M(\lambda)]=0$ leads to a
recurrent relations between $V_k(x,t)=V_{k}^{\rm f}+V_{k}^{\rm d}$
\begin{eqnarray}\label{eq:RecLax}
&& V_{k+1}^{\rm f}(x,t)\equiv \pi_J(V_{k+1}) =\Lambda_\pm
V_k^{\rm f}(x,t)- \ad_{J}^{-1} [C_k,Q(x,t)], \\
&& V_{k}^{\rm d}(x,t)\equiv (\openone -\pi_J)(V_{k}) =C_k +
\i\int_{\pm \infty}^x \rd y\,[ Q(y,t), V_k^{\rm f}(y,t)], \qquad
k=1,...,N; \nonumber
\end{eqnarray}
where $ \pi_J=\ad_J^{-1}\circ\ad_J$ and $C_k=(\openone -\pi_J)C_k$
are block-diagonal integration constants, for details see, e.g.
\cite{AKNS,ForKu*83}. These relations are resolved by the
recursion operators (\ref{eq:Lamb}):
\begin{eqnarray}\label{eq:Lamb}
\Lambda _\pm Z &=& {\mbox{ad}_J\over 4a^2} \left( \i {\rd Z\over
\rd x } + \i \left[ Q(x) , \int_{\pm\infty }^{x} \rd y \; [Q(y),
Z(y)] \right] \right).
\end{eqnarray}
where we assume that $Z \equiv \pi_J Z \in \mathcal{M}_J$. As a
result we obtain that the class of (generically nonlocal) NLEE
solvable by the ISM have the form:
\begin{equation}\label{eq:NLEE}
\i\,\ad_{J}^{-1} {\partial Q \over \partial t } +
\sum_{k=0}^{N}\Lambda_\pm^{N-k}
\left[C_k,\ad_{J}^{-1}Q(x,t)\right]=0, \qquad f(\lambda ) =\left(
\begin{array}{cc} f_+(\lambda ) & 0 \\ 0& f_-(\lambda
)\end{array}\right),
\end{equation}
where $f(\lambda ) =\sum_{k=0}^{N}C_k\lambda ^{N-k} $ determines
their dispersion law. The NLEE (\ref{eq:NLEE}) become local if
$f(\lambda )=f_0(\lambda )J $, where $f_0(\lambda ) $ is a scalar
function. In particular, if $f(\lambda )=-2\lambda ^2J $ we get
the MNLS eq. (\ref{eq:MNLSeq}).


\subsection{Basic physical example:
 ${\bf C.I}$ type symmetric space ${\rm Sp(2p)}/{\rm U(p)}$.}\label{ssec:2.3a}

We choose $\mathfrak{g} \equiv {\bf C}_2 =\mathfrak{sp}(4)$
algebra; it has 2 simple roots, namely $\alpha_1=e_1-e_2$,
$\alpha_2=2e_2$. We fix up the Cartan element
\begin{eqnarray}\label{eq:C2.0}
J = \mbox{diag}\, (a,a,-a,-a), \qquad J^2=a^2\openone .
\end{eqnarray}
Then the corresponding potential $Q(x,t)$ (\ref{eq:KF14.4}) takes the
form:
\begin{eqnarray}\label{eq:C2.1}
Q(x,t)=\left( \begin{array}{cccccc} 0 & 0 & q_{12}& q_{1} \\
0 & 0 & q_{2} &  -q_{12} \\ p_{12}& p_{2}& 0 & 0 \\
p_{1} &  -p_{12}& 0 & 0 \\ \end{array} \right)
\end{eqnarray}
Imposing the involution $p_k=q_k^* $, $k=1,2 $ and
$p_{12}=q_{12}^* $ we obtain the following 3 component MNLS system
for the independent fields $q_{12}(x,t)$, $q_{1}(x,t)$ and
$q_{2}(x,t)$:
\begin{eqnarray}\label{eq:C4.5}
&& \i a {\partial q_{12}\over \partial t} + {\partial^2
q_{12}\over
\partial x^2} + 2 q_{12}(|q_{12}|^2 + |q_{1}|^2 +|q_{2}|^2)
-2 q_{1} q_{2} q_{12}^{*}=0,
 \nonumber\\
 && \i a {\partial q_{1}\over \partial t} + {\partial^2 q_{1}\over
\partial x^2} + 2q_{1}(|q_{1}|^2 +2 |q_{12}|^2) -2q_{12}^2
q_{2}^{*}=0,\\
&& \i a {\partial q_{2}\over \partial t} + {\partial^2 q_{2}\over
\partial x^2} + 2 q_{2}(|q_{2}|^2 +2 |q_{12}|^2) -2q_{12}^2
q_{1}^{*} =0, \nonumber
\end{eqnarray}
If we identify the physical quantities of the system (\ref{eq:1}) with
\begin{eqnarray}
\Phi_{0}=q_{12},\qquad \Phi_{1}=-q_{1},\qquad \Phi_{-1}=q_{2}.
\end{eqnarray}
then eq. (\ref{eq:C4.5}) will coincide with (\ref{eq:1}). The
system can be written in a Hamiltonian form by introducing the
Poisson brackets:
\begin{equation}\label{eq:PB_C2}
\{q_j(x), p_k(y)\} = 2 \i \delta _{kj}\delta (x-y), \qquad \{
q_{12}(x), p_{12}(y)\} = \i \delta (x-y),
\end{equation}
and the Hamiltonian:
\begin{eqnarray}\label{eq:H_C2}
H&=&H_{\rm kin}+H_{\rm int},\nonumber \\ H_{\rm kin}&=&{1\over
a}\int_{-\infty}^{\infty}\rd x\, \left({\partial \Phi_{0} \over
\partial x}{\partial \Phi_{0}^{*} \over
\partial x} +\frac{1}{2}\left({\partial \Phi_{1} \over \partial
x}{\partial \Phi_{1}^{*} \over
\partial x}+{\partial \Phi_{-1} \over \partial
x}{\partial \Phi_{-1}^{*} \over
\partial x}\right)\right),\\
H_{\rm int}&=&-{1\over 2 a}\int_{-\infty}^{\infty}\rd x\, \left(
(|\Phi_{0}|^2 +|\Phi_{1}|^2 )^2+(|\Phi_{0}|^2 +|\Phi_{-1}|^2 )^2
\right) \nonumber\\
&-& \frac{1}{a}\int_{-\infty}^{\infty}\rd x\,\left(
|\Phi_{0}\Phi_{-1}^{*} +\Phi_{1}\Phi_{0}^{*}|^2 \right) .\nonumber
\end{eqnarray}
The soliton solutions of the $\mathfrak{sp}(4)$ MNLS (\ref{eq:1})
were derived independently in \cite{IMW04,I04}.

\subsection{ ${\bf D.III}$ type symmetric space}\label{ssec:2.3}

We choose $\mathfrak{g} \equiv {\bf D}_6\simeq \mathfrak{so}(12)$; it has
6 simple roots, namely $\alpha_1=e_1-e_2$, $\alpha_2=e_2-e_3$,
$\alpha_3=e_3-e_4$, $\alpha_4=e_4-e_5$, $\alpha_5=e_5-e_6$ and
$\alpha_6=e_5+e_6$. We fix up the Cartan element

\begin{eqnarray}\label{eq:D6.0}
J = \mbox{diag}\, (a,a,a,a,a,a,-a,-a,-a,-a,-a,-a), \qquad
J^2=a^2\openone
\end{eqnarray}
which means that the subset $\theta _+=\{e_i+e_j\} $ with $1\leq
i<j\leq 6 $. Then the corresponding potential $Q(x,t)$
(\ref{eq:KF14.4}) takes the form:
\begin{equation}\label{eq:D6.1}
Q=\left( \begin{array}{ccccccccccccc} 0 & 0 & 0 & 0 & 0 & 0 &
q_{16}& q_{15} & q_{14} & q_{13} & q_{12} & 0 \\
0 & 0 & 0 & 0 & 0 & 0& q_{26} & q_{25} & q_{24} &q_{23} & 0 & q_{12} \\
0 & 0 & 0 & 0 & 0 & 0 & q_{36} & q_{35} & q_{34} & 0 & q_{23} & -q_{13} \\
0 & 0 & 0 & 0 & 0 & 0 & q_{46} & q_{45} & 0 & q_{34} & -q_{24} & q_{14} \\
0 & 0 & 0 & 0 & 0 & 0 & q_{56} & 0 & q_{45} & -q_{35} & q_{25} & -q_{15} \\
0 & 0 & 0 & 0 & 0 & 0 & 0 & q_{56} & -q_{46} & q_{36} & -q_{26} & q_{16} \\
p_{16}&p_{26} & p_{36} & p_{46} & p_{56} & 0 & 0 & 0 & 0 & 0 & 0 & \\
p_{15} & p_{25} & p_{35} &p_{45} & 0 & p_{56} & 0 & 0 & 0 & 0 & 0 & 0 \\
p_{14} & p_{24} & p_{34} & 0 & p_{45} & -p_{46} & 0 & 0 & 0 & 0 & 0 & 0\\
p_{13} & p_{23} & 0 & p_{34} & -p_{35} & p_{36} & 0 & 0 & 0 & 0 & 0 & 0\\
p_{12} & 0 & p_{23} & -p_{24} & p_{25} & -p_{26} & 0 & 0 & 0 & 0 & 0 & 0\\
0 & p_{12} & -p_{13} & p_{14} & -p_{15} & p_{16} & 0 & 0 & 0 & 0 & 0 & 0\\
\end{array} \right).
\end{equation}
Here by $q_{ij}(x,t)$ and $p_{ij}(x,t)$ where $i,j $ belong to the
set of indices ${\mathcal J}=\{(ij)\,|\, 1\leq i<j\leq 6\}$ we
denote the coefficients of the generators $E_{\alpha}$ and
$E_{-\alpha}$ with $\alpha =e_i+e_j$. Then the generic NLEE
(\ref{eq:MNLSeq}) becomes a system of 30 equations. Imposing the
natural involution $Q=Q^\dag $, i.e. $p_{ij}=q_{ij}^* $  we obtain
a MNLS for the 15 independent functions $q_{ij}(x,t) $.

Similarly we consider also the ${\bf D}_5\simeq \mathfrak{so}(10)$
algebra. The corresponding $Q(x,t) $  will be a $10\times 10 $
matrix-valued function which can be obtained from the equation for
(\ref{eq:D6.1}) by removing the first and the last rows and
columns. The generic NLEE will be a system of 20 equations, the
involution $Q=Q^\dag $ will reduce them to a 10-component MNLS.

\section{Spectral data and generalized exponents}\label{sec:3}

Here we will start with a brief sketch of the direct scattering
problem for (\ref{eq:Lax-MNLS}). It is based on the Jost solutions
\cite{FaTa,ZMNP} defined by their asymptotics
\[
\lim_{x \to \infty }\psi (x,t,\lambda )\re^{\ri \lambda
Jx}=\openone , \qquad \lim_{x \to -\infty }\phi (x,t,\lambda
)\re^{\ri \lambda Jx}=\openone,
\]
and the scattering matrix $T(t,\lambda )=(\psi (x,t,\lambda
))^{-1}\phi (x,t,\lambda ) $ and its inverse $\hat{T}(\lambda ,t)
$:
\begin{eqnarray}\label{eq:ScM}
T(t,\lambda )= \left(\begin{array}{cc} a^+(t,\lambda) & -b^-(t,\lambda)\\
b^+(t,\lambda) & a^-(t,\lambda)\\ \end{array} \right), \quad
\hat{T}(t,\lambda )= \left(\begin{array}{cc} c^-(t,\lambda) &
d^-(t,\lambda)\\ -d^+(t,\lambda) & c^+(t,\lambda)\\ \end{array}
\right),
\end{eqnarray}
where $a^\pm(t,\lambda)$ and $b^\pm(t,\lambda)$ are $r\times r $
block matrices. The blocks $a^\pm$, $b^\pm$, $c^\pm$ and $d^\pm$
satisfy a number of relations coming from the fact that
$T(\lambda)\hat{T(\lambda)}=\openone $; for example
\begin{equation}\label{eq:T-hT}
a^+(\lambda) c^-(\lambda) + b^-(\lambda)d^+(\lambda)=\openone,
\qquad a^+(\lambda) d^-(\lambda) - b^-(\lambda)c^+(\lambda)=0,
\end{equation}
etc.

The fundamental analytic solutions (FAS) $\chi^{\pm} (x,t,\lambda
) $ of $L(\lambda ) $ are analytic functions of $\lambda $ for
$\mbox{Im}\,\lambda \gtrless 0$ and are related to the Jost
solutions by:
\begin{equation}\label{eq:FAS_J}
\chi ^\pm(x,t,\lambda ) = \phi (x,t,\lambda ) S_{J}^{\pm}
(t,\lambda ) = \psi (x,t,\lambda ) T_{J}^{\mp}(t,\lambda ).
\end{equation}
Here $S_{J}^{\pm} $, $T_{J}^{\pm} $ upper- and lower-
block-triangular matrices:
\begin{eqnarray}\label{eq:S_Jpm}
S_J^+(t,\lambda )= \left(\begin{array}{cc} \openone & d^-(t,\lambda)\\
0 & c^+(t,\lambda)\\ \end{array} \right), \qquad
S_J^-(t,\lambda )= \left(\begin{array}{cc} c^-(t,\lambda) & 0 \\
-d^+(t,\lambda) & \openone \\ \end{array} \right),\\
T_J^+(t,\lambda )= \left(\begin{array}{cc} \openone & -b^-(t,
\lambda)\\ 0 & a^-(t,\lambda)\\ \end{array} \right), \qquad
T_J^-(t,\lambda )= \left(\begin{array}{cc} a^+(t,\lambda) & 0 \\
b^+(t,\lambda) & \openone \\ \end{array} \right),
\end{eqnarray}
satisfy $T_J^\pm(t,\lambda ) \hat{S}_J^\pm (t,\lambda
)=T(t,\lambda ) $ and can be viewed as the factors of a
generalized Gauss decompositions of $T(t,\lambda ) $
\cite{VSG*94}. If $Q(x,t) $ evolves according to (\ref{eq:NLEE})
then
\begin{equation}\label{eq:evol}
\i {\rd b^{\pm} \over \rd t } +f_{\mp} (\lambda )
b^{\pm}(t,\lambda ) - b^{\pm}(t,\lambda ) f_{\pm}(\lambda )=0,
\quad \i {\rd a^{\pm} \over \rd t } + [f_{\pm} (\lambda ) ,
a^{\pm} (t,\lambda )] =0.
\end{equation}
On the real axis in the complex $\lambda$-plane both FAS $\chi^\pm
(x,t,\lambda)$ are linearly dependent:
\begin{equation}\label{eq:RHP}
\chi^+(x,t,\lambda)=\chi^-(x,t,\lambda)G_0(t,\lambda)
\end{equation}
and $G_0(t,\lambda)$ can be considered as a minimal set of
scattering data in the case of absence of discrete eigenvalues for
the Lax operator (\ref{eq:Lax-MNLS}), see \cite{VSG-pres}.

The mapping between the potential of the Lax operator and the
scattering data is based on the Wronskian relations \cite{CaDe}.
As an example of one we write down:
\[
b^+_{\alpha }(t,\lambda)= {\i\over 2}\int_{-\infty}^{\infty}\rd
x\, \langle
[Q(x,t),J]\chi^+(x,t,\lambda)E_{\alpha}(\chi^+(x,t,\lambda))^{-1}\rangle,
\]
where $E_{\alpha }$ is the root vector corresponding to the root
$\alpha \in\theta _+ $. Thus the ``squared'' solutions that
appeared first in \cite{AKNS,K} were later generalized in
\cite{G*86,VSG*94} to Lax operators of the type
(\ref{eq:Lax-MNLS}) as follows:
\begin{equation}\label{eq:sq_s}
e_{\alpha }^{\pm}(x,t,\lambda ) = \pi_J \left( \chi
^\pm(x,t,\lambda ) E_{\alpha }(\chi ^\pm(x,t,\lambda ))^{-1}
\right) .
\end{equation}
They can also be viewed as natural generalizations of the usual
exponentials and their completeness relations in $\mathcal{M}_J $
\cite{G*86,VSG*94} provide us the spectral decompositions of the
recursion operators $\Lambda _\pm $ for which $e_{\pm\alpha
}^{\pm}(x,\lambda ) $ are eigenfunctions:
\begin{eqnarray}\label{eq:Lam}
\Lambda _+ e_{\mp \alpha }^{\pm}(x,\lambda ) = \lambda e_{\mp
\alpha }^{\pm} (x,\lambda ), \qquad \Lambda _- e_{\pm \alpha
}^{\pm} (x,\lambda ) = \lambda e_{\pm \alpha }^{\pm} (x,\lambda ),
\qquad \alpha \in \theta _+.
\end{eqnarray}

The (generating) recursion operators $\Lambda _\pm $ appeared
first in the AKNS-approach \cite{AKNS} as a tool to generate the
class of all $M$-operators as well as the NLEE related to the
given Lax operator. Next I. M. Gel'fand and L. A. Dickey
\cite{grah:wildgelf} discovered that the class of these
$M$-operators is contained in the diagonal of the resolvent of
$L$. The kernel of the resolvent of $L$ can be explicitly defined
in terms of the fundamental analytic solutions $\chi ^\pm
(x,\lambda)$ of (\ref{eq:Lax-MNLS}), see \cite{LMP,G*86,VSG*94}.

\section{Hamiltonian properties of the MNLS models}\label{sec:4a}

It is well known that the MNLS equations possess hierarchies of
Hamiltonian structures. The phase space $\mathcal{M}_J $ of the
MNLS equations is the co-adjoint orbit of the $\mathfrak{g} \simeq
{\bf D}_r$ determined by $J $; in addition we assume that  the
matrix elements of $Q(x,t) $ are smooth functions tending to zero
fast enough for $|x|\to\infty  $.

On the ${\bf D.III} $-type symmetric spaces the Hamiltonian of
(\ref{eq:MNLSeq}) is given by:
\begin{equation}\label{eq:MNLSham1}
H^{(0)}_{ \mathrm{MNLS}} = a\int_{-\infty}^\infty \rd x\, \left\{
2\langle \ad_J^{-1}Q_x, \ad_J^{-1}Q_x \rangle +{1\over 2}\langle [
\ad_J^{-1}Q,  Q ], [ \ad_J^{-1}Q, Q ] \rangle \right\}.
\end{equation}
Direct calculation shows that this Hamiltonian is proportional to
the third coefficient $I_3$ of the expansion of the generating
functional of the principal series of motion $\ln \det \a^{+}$
with respect to $\lambda$:
\[
\ln \det\,a^+ (t, \lambda)=\sum_{k=1}^\infty I_{k}\lambda^{-k},
\]
i.e. $H^{(0)}_{ \mathrm{MNLS}}=8\i I_3$. The equation of motion
(\ref{eq:MNLSeq}) that $Q$ satisfies is generated by the canonical
symplectic structure:
\[
\Omega _{\rm MNLS}^{(0)} = -2 \i a\int_{-\infty }^{\infty } \rd x
\left\langle \delta Q(x,t) \wedgecomma \ad_J^{-1}\delta
Q(x,t)\right\rangle.
\]

The hierarchies of symplectic structures defined on $\mathcal{M}_J
$ are generated by the corresponding recursion operators
$\Lambda_{\pm}$ (\ref{eq:Lamb}) and are given by the following
families of compatible two-forms:
\begin{eqnarray*}
\Omega _{\rm MNLS}^{(k)} = -2 \i a\int_{-\infty }^{\infty } \rd x
\left\langle \delta Q(x,t) \wedgecomma \Lambda ^k \ad_J^{-1}\delta
Q(x,t)\right\rangle,\qquad
\Lambda=\frac12\left(\Lambda_{+}+\Lambda_{-}\right).
\end{eqnarray*}
The corresponding Hamiltonians for the higher MNLS are $H^{(k)}_{
\mathrm{MNLS}}=8\i I_{3+k}$.

For $f(\lambda )=-2\lambda ^2J $ eq. (\ref{eq:evol}) gives $\rd
a^\pm/\rd t=0 $ and  $a^{\pm} (\lambda)$ can be viewed as
generating functionals of integrals of motion whose number $r^2$
is larger than the rank $r$ of $\mathfrak{g}$.  This is obviously
due to the degeneracy of the dispersion law. For generic
$f(\lambda ) $ from (\ref{eq:evol}) there follows that only
functions of the eigenvalues of $a^\pm(\lambda ) $ will be
conserved. Indeed, it follows from the classical $R$-matrix
approach \cite{FaTa}. One of the definitions of the classical
$R$-matrix is based on the Lax representation of the MNLS
(\ref{eq:Lax-MNLS}). In particular, if $U(x,\lambda)$ has the form
\[
U(x,\lambda) = Q(x,t)-\lambda J
\]
and the matrix elements of $Q(x,\lambda)$ satisfy the Poisson
brackets $\{\cdot,\cdot\}$:
\begin{equation}\label{eq:Pois}
\{p_{\alpha_{2}}(x,t),q_{\alpha_{1}}(y,t) \}=\i
\frac{(\alpha_{1},\alpha_{1})}{2}\;\delta_{\alpha_{1} \alpha_{2}}
\delta(x-y)
\end{equation}
then the classical $R$--matrix can be defined through the relation
\begin{equation}\label{eq:UPois2}
\{U(x,\lambda)\otimescomma U(y,\mu)\}=[R(x-y), U(x,\lambda)\otimes
\openone + \openone \otimes U(y,\mu)]\delta(x-y)
\end{equation}
Our system of equations (\ref{eq:UPois2}) allows $R$--matrix given
by\cite{ForKu*83}:
\begin{equation}\label{eq:R}
    R(\lambda-\mu)=\frac{\i}{2} {1\over \lambda-\mu}P,
\end{equation}
where
\begin{equation}
P= \left( \sum_{k=1}^{r} h_k \otimes h_k + \sum_{\alpha \in \Delta
}  {E_{\alpha } \otimes E_{-\alpha} \over \langle E_\alpha
,E_{-\alpha }\rangle} \right).
\end{equation}
Here $h_k$ are introduced in (\ref{eq:31.2}) and are properly
normalized to $\langle h_i, h_k\rangle =\delta_{ik}$. $P$ is the
second Casimir endomorphism of the algebra and posses  special
properties concerning its action onto the matrices of the
corresponding group:
\[
P(A\otimes B)=(B\otimes A)P,
\]
Using these properties of the $P$--matrix and the commutation
relations of the algebra (\ref{eq:31.2}) we obtain
\begin{eqnarray*}
&&[P, Q(x)\otimes \openone + \openone \otimes Q(x)]=0 \\
&&[P,\lambda J\otimes \openone+\mu \openone \otimes
J]=2(\lambda-\mu)\{U(x,\lambda)\otimescomma U(x,\mu)\}
\end{eqnarray*}
These relations are true for any $Q(x)$ taking value in the
algebra and the relation (\ref{eq:UPois2}) seems most natural and
its r.h.side does not contain $Q(x,t)$.

Let us now show, that the classical $R$--matrix is a very
effective tool for calculating the Poisson brackets between the
matrix elements of $T (\lambda )$. It will be more convenient here
to consider periodic boundary conditions on the interval $[-L,L]
$, i.e. $Q(x-L)=Q(x+L) $ and to introduce the fundamental solution
$T(x,y,\lambda ) $ \cite{FaTa}:
\begin{equation}\label{eq:eqT}
\i {\rd T(x,y,\lambda )  \over \rd x } + U(x,\lambda
)T(x,y,\lambda ) =0, \qquad T(x,x,\lambda )=\openone .
\end{equation}

Skipping the details we just formulate the following relation for
the Poisson brackets between the matrix elements of $T(x,y,\lambda
) $:
\begin{equation}\label{eq:L5.12}
\left\{ T(x,y,\lambda ) \otimescomma  T(x,y,\mu )\right\} = \left[
R(\lambda -\mu ), T(x,y,\lambda )\otimes T(x,y,\mu )\right]
\end{equation}

The corresponding monodromy matrix $T_L(\lambda ) $ describes the
transition from $-L $ to $L $ and $T_L(\lambda ) =T(-L,L,\lambda
)$. The Poisson brackets between the matrix elements of
$T_L(\lambda )$ follow directly from eq. (\ref{eq:L5.12}) and are
given by:
\begin{equation}\label{eq:L5.20}
\left\{T_L (\lambda ) \otimescomma T_L(\mu )\right\} =
\left[R(\lambda -\mu ), T_L(\lambda ) \otimes T_L (\mu )\right]  .
\end{equation}
We could also write the Poisson brackets between the matrix
elements of the inverse of the monodromy matrix $\hat{T}_L(\lambda
)$:
\begin{equation}\label{eq:L5.20_inverse}
\left\{\hat{T}_L (\lambda ) \otimescomma \hat{T}_L(\mu )\right\} =
\left[\hat{T}_L(\lambda ) \otimes \hat{T}_L (\mu ),R(\lambda -\mu
)\right] .
\end{equation}

An elementary consequence of this result is the involutivity of
the integrals of motion $I_{L,k}$ and $J_{L,k} $ from the
principal series which appear in the expansions of:
\begin{eqnarray}\label{eq:ln-det}
\ln \det \a_L^+(\lambda ) = \sum_{k=1}^{\infty } I_{L,k}\lambda
^{-k}, \qquad  -\ln \det \c_L^-(\lambda ) = \sum_{k=1}^{\infty }
J_{L,k}\lambda
^{-k}, \\
\ln \det \c_L^+(\lambda ) = \sum_{k=1}^{\infty } J_{L,k}\lambda
^{-k}, \qquad  -\ln \det \a_L^-(\lambda ) = \sum_{k=1}^{\infty }
I_{L,k}\lambda ^{-k},
\end{eqnarray}
As a result of the reduction conditions (\ref{eq:a-c.I.red}) that
we impose, the generating functional of the principal series of
integrals of motion is only one, i.e. $\ln \det \a_L^+(\lambda )$.

Another important property of the integrals $I_{L,k}$ and $J_{L,k}
$ is their locality, i.e. their densities depend only on $Q $ and
its $x $-derivatives.

The simplest consequence of the  relations (\ref{eq:L5.20}) and
(\ref{eq:L5.20_inverse}) is the involutivity of $I_{L,k}$ and
$J_{L,k} $. Indeed, taking the trace of both sides of
(\ref{eq:L5.20}) and (\ref{eq:L5.20_inverse}) shows that $\{ \tr
T_L(\lambda ),\tr T_L(\mu )\}=0 $ and $\{ \tr \hat{T}_L(\lambda
),\tr \hat{T}_L(\mu )\}=0 $.  We can also multiply both sides of
(\ref{eq:L5.20}) and (\ref{eq:L5.20_inverse}) by $C\otimes C$ and
then take the trace. This proves:
\begin{equation}\label{eq:L5.23}
\left\{\tr T_L (\lambda )C, \tr T_L (\mu )C\right\} = 0,\qquad
\left\{ \tr \hat{T}_L(\lambda )C,\tr \hat{T}_L(\mu )C\right\}=0.
\end{equation}

In particular, for $C = \openone +J$  and $C = \openone -J$  we
get the involutivity of:
\begin{equation}\label{eq:L5.24}
\begin{split}
\left\{ \tr \a^{\pm}_L(\lambda ),\tr \a^{\pm}_L(\mu )\right\} =0,
\qquad
\left\{\tr \a^+_L(\lambda ),\tr \a^-_L(\mu )\right\} = 0 ,\\
\left\{ \tr \c^{\pm}_L(\lambda ),\tr \c^{\pm}_L(\mu )\right\} =0,
\qquad \left\{\tr \c^+_L(\lambda ),\tr \c^-_L(\mu )\right\} = 0 ,
\end{split}
\end{equation}

Equations (\ref{eq:L5.20}) and (\ref{eq:L5.20_inverse}) were
derived for the typical representation $V^{(1)} $ of the
corresponding group ${\rm G}$, but they hold true also for
any other finite-dimensional representation of ${\rm G} $.
Let us denote by $V^{(r)}$ the $r $-th fundamental representation
of ${\rm G} $; then the element $T_L(\lambda ) $ will have
representation in $V^{(r)} $, see \cite{Helg}. In particular, if
we consider equations (\ref{eq:L5.20}) and
(\ref{eq:L5.20_inverse}) in the representation $V^{(r)} $ and
sandwich them between the highest and lowest weight vectors in
$V^{(r)} $ we get:
\begin{equation}\label{eq:det-a}
\{ \det \a^+_L(\lambda ), \det \a^+_L(\mu )\} =0, \qquad \{ \det
\c^-_L(\lambda ), \det \c^-_L(\mu )\} =0.
\end{equation}

Since eq. (\ref{eq:det-a})  hold true for all values of $\lambda $
and $\mu  $ we can insert into them the expansions
(\ref{eq:ln-det}) with the result:
\begin{equation}\label{eq:I_L}
\{ I_{L,k}, I_{L,p}\} =0,\qquad \{ J_{L,k}, J_{L,p}\} =0,\qquad
k,p= 1,2,\dots.
\end{equation}

Lets go back to our basic physical example from Section
\ref{ssec:2.3a} associated with the algebra $\mathfrak{g} \equiv
\mathfrak{sp}(4)$ and consider the monodromy matrix
$T_{L}(\lambda)$ in the typical representation. The matrix
elements of the diagonal blocks of $T_{L\;ab}(\lambda )$ for $1
\leq a \leq 2$ and $1 \leq b \leq 2$ are denoted by
$T_{L\;ab}(\lambda )=\a^+_{ab}$ according to (\ref{eq:ScM}). Now
Poisson brackets between the scattering data $\a^{+}_L (\lambda)$
and $\a^{+}_L (\mu)$ are obtained in a straightforwar way from:
(\ref{eq:L5.20}):
\begin{equation}\label{}
\begin{split}
&\left\{\a^{+}_{L\;11}(\lambda ), \a^{+}_{L\;12} (\mu )\right\}=
\frac{\i}{2(
\lambda-\mu)}\left(\a^{+}_{L\;11}(\lambda)\a^{+}_{L\;12}(\mu)-\a^{+}_{L\;12}(\lambda)
\a^{+}_{L\;11}(\mu) \right)\\
&\left\{\a^{+}_{L\;11}(\lambda ), \a^{+}_{L\;21} (\mu )\right\}=
\frac{\i}{2(
\lambda-\mu)}\left(\a^{+}_{L\;21}(\lambda)\a^{+}_{L\;11}(\mu)-\a^{+}_{L\;11}(\lambda)
\a^{+}_{L\;21}(\mu) \right)\\
&\left\{\a^{+}_{L\;11}(\lambda ), \a^{+}_{L\;22} (\mu )\right\}=
\frac{\i}{2(
\lambda-\mu)}\left(\a^{+}_{L\;21}(\lambda)\a^{+}_{L\;12}(\mu)-\a^{+}_{L\;12}(\lambda)
\a^{+}_{L\;21}(\mu) \right)\\
&\left\{\a^{+}_{L\;22}(\lambda ), \a^{+}_{L\;12} (\mu )\right\}=
\frac{\i}{2(
\lambda-\mu)}\left(\a^{+}_{L\;12}(\lambda)\a^{+}_{L\;22}(\mu)-\a^{+}_{L\;22}(\lambda)
\a^{+}_{L\;12}(\mu) \right)\\
&\left\{\a^{+}_{L\;22}(\lambda ), \a^{+}_{L\;21} (\mu )\right\}=
\frac{\i}{2(
\lambda-\mu)}\left(\a^{+}_{L\;22}(\lambda)\a^{+}_{L\;21}(\mu)-\a^{+}_{L\;21}(\lambda)
\a^{+}_{L\;22}(\mu) \right)\\
&\left\{\a^{+}_{L\;21}(\lambda ), \a^{+}_{L\;12} (\mu )\right\}=
\frac{\i}{2(
\lambda-\mu)}\left(\a^{+}_{L\;11}(\lambda)\a^{+}_{L\;22}(\mu)-\a^{+}_{L\;22}(\lambda)
\a^{+}_{L\;11}(\mu) \right)\\
&\left\{\a^{+}_{L\;11}(\lambda ), \a^{+}_{L\;11} (\mu )\right\}=
0,\qquad \left\{\a^{+}_{L\;22}(\lambda ), \a^{+}_{L\;22} (\mu
)\right\}= 0\\
&\left\{\a^{+}_{L\;12}(\lambda ), \a^{+}_{L\;12} (\mu )\right\}=
0,\qquad \left\{\a^{+}_{L\;21}(\lambda ), \a^{+}_{L\;21} (\mu
)\right\}= 0.\\
\end{split}
\end{equation}
Now that we know the Poisson brackets between the matrix elements
of $\a^{+}$ it is not difficult to extend the above result over
its invariants:
\begin{eqnarray*}\label{}
&&\{\tr \a^{+}_{L}(\lambda), \a^{+}_{L\;ij}(\mu)\}={\i \over
2}\frac{1}{ \lambda-\mu}[
\a^{+}_L(\lambda),\a^{+}_L(\mu)]_{ij},\quad{\rm
for}\quad i=1,2\;j=1,2\\
&&\{\det \a^{+}_L(\lambda), \a^{+}_{L\;ij}(\mu)\}=0,\quad{\rm
for}\quad i=1,2\;j=1,2
\end{eqnarray*}
This somewhat more concrete analysis allows one to see that only
functions of the eigenvalues of $\a^{+}_{L}(\lambda ) $  produce
integrals of motion in involution.

We are able to transfer these results also for the case of
potentials belonging to the co-adjoint orbit of the $\mathfrak{g}
\simeq {\bf D}_r$ determined by $J $ and with zero boundary
conditions
\begin{eqnarray}\label{eq:T-T}
\{T(\lambda,t)\otimescomma T(\mu,t)\} =[R(\lambda-\mu),
T(\lambda,t)\otimes T(\mu,t)].
\end{eqnarray}
Here $T(\lambda,t)$ is the scattering matrix, obtained after
taking the limit $L \to \infty $ in the corresponding monodromy
matrix $T_L(\lambda )$. Indeed, let us multiply (\ref{eq:L5.12})
by $E(y,\lambda ) \otimes E(y,\mu )$ on the right and by $E^{-1}
(x,\lambda ) \otimes E^{-1} (x,\mu )$  on the left, where
$E(x,\lambda)=\exp (-\i\lambda Jx)$ and take the limit for $x \to
\infty $, $y \to -\infty $. Since:
\begin{equation}\label{eq:L5.26}
\lim_{x\to\pm \infty } {{\rm e}^{\i x(\lambda -\mu )} \over
\lambda - \mu } =\pm \i \pi  \delta (\lambda  - \mu ),
\end{equation}
we get:
\begin{eqnarray}\label{eq:L5.27}
\begin{split}
&\left\{T (\lambda ) \otimescomma T (\mu )\right\} = R_+ (\lambda
- \mu )T (\lambda ) \otimes T (\mu ) -
T (\lambda ) \otimes T (\mu ) R_- (\lambda -\mu ) , \\
&R_\pm (\lambda -\mu )={1\over 2(\lambda  - \mu ) }
\left(\sum_{k=1}^{r} h_k \otimes h_k +\Pi_{0J} \right) \pm \i\pi
\delta (\lambda -\mu )\Pi_{1J},
\end{split}
\end{eqnarray}
where $\Pi_{0J}$ and $\Pi_{1J}$ are defined as follows
\begin{eqnarray}\label{eq:5.23'}
\begin{split}
&\Pi_{0J} =\sum_{\alpha \in \theta _0}\left(
 E_{\alpha}\otimes E_{-\alpha}+ E_{-\alpha}\otimes E_{\alpha}\right),\\
&\Pi_{1J} =\sum_{\alpha \in \theta _+}\left(
 E_{\alpha}\otimes E_{-\alpha}- E_{-\alpha}\otimes E_{\alpha}\right).
\end{split}
\end{eqnarray}
Analogously we prove that: \\
i) there are integrals $I_k=\lim_{L\to\infty }I_{L,k} $ and
$J_p=\lim_{L\to\infty }J_{L,p} $ that are in involution, i.e.:
\[ \{I_k, I^{\pm}_p\}  =  \{J_k, J_p\} = 0, \]
for some positive values of $k $ and $p $; ii) the principal
series of integrals of motion $I_{k}$, generated by $\ln
\det\,a^\pm (t, \lambda)$, i.e. the eigenvalues of $\a^\pm(\lambda
) $ and $\c^\pm(\lambda ) $ produce integrals of motion in
involution. So for MNLS we have extra integrals of motion that are
not all in involution. Indeed, lets denote the matrix elements of
the scattering matrix $T(\lambda,t)$ according to (\ref{eq:ScM})
and multiply (\ref{eq:L5.27}) by $E_{ab}\otimes E_{cd}$ on the
right, where $(E_{ab})_{ij}=\delta_{ai}\delta_{bj}$ and take the
trace of the elements in the first and in the second position of
the tensor product. Thus we obtain the Poisson brackets between
the scattering data:
\begin{equation}\label{eq:el-Pois}
\left\{T _{ba}(\lambda ), T_{dc} (\mu )\right\} =\tr \left(  T
(\lambda )\otimes T (\mu ) \left(E_{ab} \otimes E_{dc} R_+ - R_-
E_{ab} \otimes E_{dc}\right)\right)
\end{equation}
Lets list the relations contained in the above equality and
concerning the block-diagonal portions of the scattering matrix
$\a^{+}(\lambda)$ and $\a^{-}(\lambda)$
\begin{equation}\label{eq:ScM-Pois}
\left\{\a^{\pm}_{ba}(\lambda ), \a^{\pm}_{dc} (\mu )\right\}=
\frac{\i}{2(
\lambda-\mu)}\left(\a^{\pm}_{da}(\lambda)\a^{\pm}_{bc}(\mu)-\a^{\pm}_{bc}(\lambda)\a^{\pm}_{da}(\mu)
\right)\\
\end{equation}
The above result allows us to compute the Poisson brackets between
the invariants of $\a^{\pm}(\lambda)$ and their matrix elements:
\begin{equation}\label{eq:ScM-Pois2}
\begin{split}
&\left\{\tr (\a^{\pm}(\lambda))^{k},
\a^{\pm}_{ab}(\mu)\right\}={\i \over 2}\frac{k}{
\lambda-\mu}\left[ \left(\a^{\pm}(\lambda)\right)^{k},\a^{\pm}(\mu)\right]_{ab}\\
&\{\tr \ln \a^{\pm}(\lambda), \a^{\pm}_{dc}(\mu)\}=0
\end{split}
\end{equation}
This analysis allows reveals that only the eigenvalues of
$\a^\pm(\lambda ) $ and $\c^\pm(\lambda )$ produce integrals of
motion in involution.

From (\ref{eq:T-T}) it follows that the first integrals $I_{k}$
generated by $ \ln \det\,a^\pm (t, \lambda) $ are in involution.
Due to the special degenerate choice of the dispersion law
$f(\lambda)=-2\lambda^2 J$, any matrix elements of the blocks
$\a^\pm(\lambda)$ will generate integrals of motion, which however
will not be in involution, see
(\ref{eq:ScM-Pois},\ref{eq:ScM-Pois2}). The Hamiltonian for the
MNLS models is proportional to $I^{+}_{3}$, i.e. belongs to the
principal series. If we choose a generic (i.e., non-degenerate)
dispersion law then the Hamiltonian of the corresponding NLEE will
not be in involution with $I^{\pm}_{k}$. Such are the dispersion
laws for the NLEE's that allow "boomeron" and "trappon" type
solutions \cite{CD2,D1,CaDe-new}. This is the reason why their
velocities become time-dependent.

\section{Dressing factors and soliton solutions}\label{sec:4b}

The main idea of the dressing method is starting from a FAS $\chi
_{(0)}^{\pm}(x,\lambda ) $ of $L(\lambda) $ with potential
$q_{(0)} $ to construct a new singular solution $\chi
_{(1)}^{\pm}(x,\lambda ) $ of the Riemann--Hilbert Problem
(\ref{eq:RHP}) with singularities located at prescribed positions
$\lambda _1^{\pm} $. Then the new solutions
$\chi_{(1)}^{\pm}(x,\lambda ) $ will correspond to a potential
$q_{(1)} $ of $L(\lambda) $ with two discrete eigenvalues $\lambda
_1^{\pm} $. It is related to the regular one by the dressing
factors $u(x,\lambda ) $:
\begin{eqnarray}\label{eq:4.1}
\chi _{(1)}^{\pm}(x,\lambda ) = u(x,\lambda ) \chi
_{(0)}^{\pm}(x,\lambda ) u_-^{-1}(\lambda),\quad u_-(\lambda
)&=&\lim_{x \to -\infty }u(x,\lambda ).
\end{eqnarray}
If $\mathfrak{g}\simeq {\bf B}_r, {\bf D}_r$ the dressing factors
take the form \cite{1}:
\begin{equation}\label{eq:4.2}
u(x,\lambda )= \openone + (c_1(\lambda ) -1)P_1(x) +
(c_1^{-1}(\lambda ) -1) P_{-1}(x), \qquad P_{-1} =SP_1^TS^{-1},
\end{equation}
where the rank 1 projector $P_1(x) $ and the 
function $c_1(\lambda)$ are given by:
\begin{eqnarray}\label{eq:4.3}
&& P_1(x)= {|n(x)\rangle \langle m(x)|\over \langle
m(x)|n(x)\rangle },\qquad c_1(\lambda)= {\lambda - \lambda^+\over
\lambda - \lambda^-},
\nonumber\\
&& |n(x)\rangle =\chi _0^+(x,\lambda_1^+)|n_{0}\rangle, \qquad
\langle m(x)|= \langle m_{0}| \hat{\chi} _0^-(x,\lambda _1^-).
\end{eqnarray}
$|n_0\rangle $ and $\langle m_0| $ are constant vectors and
\begin{equation}
S=\sum_{k=1}^{r} (-1)^{k+1} (E_{k\bar{k}} + E_{\bar{k}k}) ,\qquad
\bar{k}= 2r+1-k. \nonumber
\end{equation}
Here $E_{kn} $ is an $2r\times 2r $ matrix whose matrix elements
are $(E_{kn})_{ij}=\delta _{ik}\delta _{nj} $. Then the
``dressed'' potential have the form
\begin{equation}\label{eq:4.4}
Q_{(1)}(x,t) = Q_{(0)}(x,t) - (\lambda _1^+ - \lambda _1^-)
[J,p(x,t)], \qquad p(x,t) = P_1(x,t) - P_{-1}(x,t).
\end{equation}
where
\begin{eqnarray}\label{eq:4.5}
p(x,t)&=& \frac{2}{ \langle m|n\rangle }\left( \sum_{k=1}^{r}
h_k(x,t) H_{e_k} + \sum_{\alpha \in \Delta _+} ( P_{\alpha }(x,t)
E_{\alpha } + P_{-\alpha }(x,t) E_{-\alpha })\right),\nonumber\\
\langle m|n\rangle &=& \sum_{k=1}^{r}(n_{0k}m_{0k}\re ^{2 a\nu_1x
+16 a \mu_1\nu_1t} +n_{0\bar{k}}m_{0\bar{k}}\re ^{-2 a\nu_1x -16 a
\mu_1\nu_1t}),\nonumber\\
h_k(x,t)&=&n_{0k}m_{0k}\re ^{2 a\nu_1x +16 a \mu_1\nu_1t}
-n_{0\bar{k}}m_{0\bar{k}}\re ^{-2 a\nu_1x -16 a
\mu_1\nu_1t},\nonumber
\end{eqnarray}
and
\begin{eqnarray}
P_{\alpha}(x,t)&=& (n_{0k}m_{0\bar{s}}-
(-1)^{s+k}n_{0s}m_{0\bar{k}})\re ^{-2\ri a\mu_1x - 8\ri a
(\mu_1^{2}-\nu_1^{2})t},\nonumber\\
P_{-\alpha}(x,t)&=& (n_{0\bar{s}}m_{0k}-
(-1)^{s+k}n_{0\bar{k}}m_{0s})\re ^{2\ri a\mu_1x + 8\ri a
(\mu_1^{2}-\nu_1^{2})t};\nonumber
\end{eqnarray}
for $\alpha=e_k+e_s$, and
\begin{eqnarray}
P_{\alpha}(x,t)&=& n_{0k}m_{0s}e ^{2 a\nu_1x + 16 a \mu_1\nu_1t} -
(-1)^{s+k}n_{0\bar{s}}m_{0\bar{k}}\re ^{-2 a\nu_1x - 16 a
\mu_1\nu_1t}, \nonumber\\
P_{-\alpha}(x,t)&=& n_{0\bar{k}}m_{0\bar{s}}\re ^{-2 a\nu_1x - 16
a \mu_1\nu_1t} - (-1)^{s+k}n_{0s}m_{0k}\re ^{2 a\nu_1x + 16 a
\mu_1\nu_1t},
\end{eqnarray}
for $ \alpha=e_k-e_s, \, k<s$.

Let us consider now the purely solitonic case, i.e. $Q_{(0)}(x)=0
$ and
\[\chi _{(0)}^{+}(x,t,\lambda_{1}^{+} )=\exp(-\i
\lambda_{1}^{+} J x-4\ri \lambda_{1}^{+2} J t). \] Thus the
1-soliton solution is
\begin{eqnarray} \label{eq:4.6}
q_{jk}&=&-{2\i a\nu_{1} \left( n_{0j} m_{0\bar{k}}
-(-1)^{j+k}n_{0k}m_{0\bar{j}} \right)e^{-2\i a \mu_{1}x-8\i a
(\mu_{1}^2-\nu_{1}^2) t } \over \sqrt{\varphi_1\varphi_2
}\mbox{ch}\left(2a\nu_{1} x +16a\nu_{1} \mu_{1}
t+\frac{1}{2}\mbox{ln}\frac{\varphi_{1}}{\varphi_{2}}\right)},
\end{eqnarray}
where $(ij)\in \mathcal{J}$ and
\begin{equation}
\varphi_{1}=\sum_{j=1}^{r} n_{0j}m_{0j},\quad
\varphi_{2}=\sum_{j=1}^{r} n_{0\bar{j}}m_{0\bar{j}},\quad
\lambda_{1}^{\pm}=\mu_{1}\pm \i\nu_{1}.
\end{equation}

\section{New reductions of MNLS equations}\label{sec:5}

\subsection{The reduction group}\label{ssec:5.1}

The reduction group $G_R $ introduced by A. V. Mikhailov \cite{2}
provides a powerful tool for constructing new integrable equations
\cite{Za*Mi,DrSok,1,vgn2,MiLo1,MiLo2} starting from known ones.
It is a finite group which preserves the Lax representation, i.e.
it ensures that the reduction constrains are automatically
compatible with the evolution. The main idea of the reduction
group is to impose an invariance condition on the Lax operators
(\ref{eq:Lax-MNLS}) and (\ref{eq:Mop}). In particular this means
that the dispersion law $f_{\rm MNLS}(\lambda)=-2\lambda^2 J $
must also be compatible with the reduction group action.

Here we consider two types of $G_R $ reductions; like in \cite{MV}
we will embed them as subgroup of $W_{\mathfrak{g}} $:
\begin{eqnarray}
\label{eq:3.1} {\rm \bf Type \,\, I:}\quad &B^{-1} U^\dag
(x,t,\lambda
^*)B = U(x,t,\lambda ), &\qquad B^{-1}J B =J, \nonumber\\
&B^{-1} V^\dag (x,t,\lambda ^*)B = V(x,t, \lambda ),&\\
\label{eq:3.3} {\rm \bf Type \,\, II:}\quad & C^{-1}
U^*(x,t,\lambda ^*)C = -U(x,t,\lambda ),& \qquad C^{-1}J C
=-J,\nonumber\\
&C^{-1} V^*(x,t,\lambda ^*)C = -V(x,t,\lambda ),&
\end{eqnarray}
where
\[
U(x,t,\lambda )=Q(x,t) - \lambda J
\]
and
\[
V(x,t,\lambda )=- [Q, \ad_J^{-1}Q] + 2\i \mbox{ad}_J^{-1}Q_x(x,t)
+2\lambda Q(x,t) -2\lambda ^2 J.
\]
The automorphisms $C $ and $B $ must be of even order.

\subsection{Example of $\bbbz_4 $-reduction.}\label{subsec:5.1}

Let us impose the following $\bbbz_4$-reduction:
\[
B^{-1}(U^{\dag}(x,t,\lambda^*))B=U(x,t,\lambda), \quad
U(x,t,\lambda)=Q(x,t)-\lambda J
\]
\[
B=w_{e_1-e_2}\cdot w_{e_2-e_3}\cdot w_{e_3-e_4},\quad
B^4=\openone,
\]
where $w_{e_i-e_j}$ are the Weyl reflection with respect to the
roots $e_i-e_j$ of the $\mathfrak{so}(10)$-algebra. Then $ B(J)=J
$ and the corresponding reduced potential $Q^{red}\in
\mathfrak{so}(10)$ takes the form:
\begin{eqnarray}\label{eq:D5.1}
Q^{\rm red}=\left( \begin{array}{ccccccccccccc}
0 & 0 & 0 & 0 & 0 & q_{15}& q_{14} & 0 & q_{12} & 0 \\
0 & 0 & 0 & 0 & 0 & q_{25} & 0 & q_{14} &0 & q_{12} \\
0 & 0 & 0 & 0 & 0 & q_{15} & q_{12} & 0 & q_{14} & 0 \\
0 & 0 & 0 & 0 & 0 & q_{25} & 0 & q_{12} & 0 & q_{14} \\
0 & 0 & 0 & 0 & 0 & 0 & q_{25} & -q_{15} & q_{25} & -q_{15} \\
q_{15}^{*}&q_{25}^{*} & q_{15}^{*} & q_{25}^{*} & 0 & 0 & 0 & 0 &
0 & 0 & \\ q_{14}^{*} & 0 & q_{12}^{*} & 0 & q_{25}^{*} & 0 & 0 & 0 & 0 &
0 \\ 0 & q_{14}^{*} & 0 & q_{12}^{*} & -q_{15}^{*} & 0 & 0 & 0 & 0 & 0 \\
q_{12}^{*} & 0 & q_{14}^{*} & 0 & q_{25}^{*} & 0 & 0 & 0 & 0 & 0 \\
0 & q_{12}^{*} & 0 & q_{14}^{*} &-q_{15}^{*} & 0 & 0 & 0 & 0 & 0 \\
\end{array} \right) .
\end{eqnarray}

Thus one derives the following 4 component MNLS system related to
${\bf D}_5$- algebra (here the independent fields are:
$q_{12}(x,t)$, $q_{14}(x,t)$, $q_{15}(x,t)$ and $q_{25}(x,t)$:
\begin{eqnarray}\label{eq:D5.5}
&& \i a {\partial q_{12}\over \partial t} + {\partial^2
q_{12}\over
\partial x^2} + 2q_{12}(|q_{12}|^2 +2 |q_{14}|^2 +|q_{15}|^2+ |q_{25}|^2)
\nonumber\\
&&+2 q_{14}(|q_{15}|^2+|q_{25}|^2)+2 q_{14}^2 q_{12}^{*}=0,
 \nonumber\\
 && \i a {\partial q_{14}\over \partial t} + {\partial^2 q_{14}\over
\partial x^2} + 2q_{14}(|q_{14}|^2 +2 |q_{12}|^2 +|q_{15}|^2+ |q_{25}|^2)
\nonumber\\
&&+2 q_{12}(|q_{15}|^2+|q_{25}|^2)+2 q_{12}^2 q_{14}^{*}=0,
 \\
 && \i a {\partial q_{15}\over \partial t} + {\partial^2 q_{15}\over
\partial x^2} + 2 q_{15}(2|q_{15}|^2 +2 |q_{25}|^2 +|q_{12}|^2 +|q_{14}|^2)
\nonumber\\
&&+2 q_{12} q_{15} q_{14}^{*}+2 q_{14} q_{15} q_{12}^{*}=0,
 \nonumber \\
&& \i a {\partial q_{25}\over \partial t} + {\partial^2
q_{25}\over
\partial x^2} + 2 q_{25}(2|q_{25}|^2 +2 |q_{15}|^2 +|q_{12}|^2 +|q_{14}|^2)
\nonumber\\
&&+2 q_{14} q_{25} q_{12}^{*}+2 q_{12} q_{25} q_{14}^{*}=0, \nonumber
\end{eqnarray}

The Hamiltonian of (\ref{eq:D5.5}) is obtained from the general
expression (\ref{eq:MNLSham1}) by imposing  the reduction
constraints:
\begin{eqnarray}\label{eq: H-D5}
H&=&H_{\rm kin}+H_{\rm int},\quad H_{\rm kin}={2\over
a}\int_{-\infty}^{\infty}\rd x\,\left(
 {\partial q_{15} \over
\partial x}{\partial  q_{15}^{*} \over
\partial x} + {\partial  q_{14}  \over \partial
x}{\partial  q_{14}^{*} \over
\partial x} \right. \nonumber \\ &+&  \left.{\partial  q_{12}  \over \partial
x}{\partial  q_{12}^{*} \over
\partial x} +{\partial q_{12} \over \partial
x}{\partial q_{12}^{*} \over \partial x}\right),\\
H_{\rm int}&=&-\frac{2}{a}\int_{-\infty}^{\infty}\rd x\,
(|q_{15}|^2+
|q_{25}|^2 + |q_{12}|^2+|q_{14}|)^2 \nonumber \\
&-& \frac{2}{a}\int_{-\infty}^{\infty}\rd x\, (|q_{15}|^2+
|q_{25}|^2 + q_{12}^*q_{14}+ q_{12}q_{14}^* )^2.
\end{eqnarray}

\subsection{Example of $\bbbz_6 $-reduction.}\label{ssec:5.2}

Let us impose the following $\bbbz_6$-reduction:
\[
B^{-1}(U^{\dag}(x,t,\lambda^*))B=U(x,t,\lambda), \quad
U(x,t,\lambda)=Q(x,t)-\lambda J
\]
\[
B=w_{e_1-e_2}\cdot w_{e_2-e_3}\cdot w_{e_3-e_4},\cdot
w_{e_4-e_5},\cdot w_{e_5-e_6},\qquad B^6=\openone,
\]
where $w_{e_i-e_j}$ are the Weyl reflection with respect to the
roots $e_i-e_j$ of the $\mathfrak{so}(12)$-algebra. Then $ B(J)=J
$ and the corresponding reduced potential $Q^{red}\in
\mathfrak{so}(12)$ takes the form:
\begin{eqnarray}
Q^{\rm red}=\left( \begin{array}{cccccc} 0 & {\bf q}\\ {\bf q}^{\dag}& 0
\end{array} \right) ,\nonumber
\end{eqnarray}
where
\begin{eqnarray}\label{seq:D6.1}
{\bf q}=\left( \begin{array}{ccccccccc} q_{12}& -q_{13} & q_{14}
& q_{13} & q_{12}
& 0 \\ -q_{13} & q_{14} & q_{13} &q_{12} & 0 & q_{12} \\
q_{14} & q_{13} & q_{12} & 0 & q_{12} & -q_{13} \\
q_{13} & q_{12} & 0 & q_{12} & -q_{13} & q_{14} \\
q_{12} & 0 & q_{12} & -q_{13} & q_{14} & q_{13} \\
0 & q_{12} & -q_{13} & q_{14} & q_{13} & q_{12} \\
\end{array} \right) .\nonumber
\end{eqnarray}

Thus one derives the following 3 component MNLS system related to
${\bf D}_6$- algebra (here the independent fields are:
$q_{12}(x,t)$, $q_{13}(x,t)$, and $q_{14}(x,t)$ :
\begin{eqnarray}\label{eq:D4.5}
&& \i a {\partial q_{12}\over \partial t} + {\partial^2
q_{12}\over
\partial x^2} - 2q_{12}(3|q_{12}|^2 +2 |q_{13}|^2 +2|q_{14}|^2)
\nonumber\\
&&+4 q_{14}(|q_{13}|^2-|q_{12}|^2)+2 q_{13}^2 q_{12}^{*}-
2q_{12}^2 q_{14}^{*}-2q_{13}^2 q_{14}^{*}-2q_{14}^2 q_{12}^{*}=0
 \nonumber\\
 && \i a {\partial q_{13}\over \partial t} + {\partial^2 q_{13}\over
\partial x^2} - 2q_{13}(3|q_{13}|^2 +2 |q_{12}|^2 +2|q_{14}|^2)
\nonumber\\
&&+2 q_{12}^2 q_{13}^{*}+ 2 q_{14}^2 q_{13}^{*}+4 q_{13} q_{12}
q_{14}^{*}+4 q_{13} q_{14} q_{12}^{*} -4 q_{12} q_{14}
q_{13}^{*}=0
 \\
 && \i a {\partial q_{14}\over \partial t} + {\partial^2 q_{14}\over
\partial x^2} - 2 q_{14}(|q_{14}|^2 +4 |q_{12}|^2 +4|q_{13}|^2)
\nonumber\\
&&+4 q_{12}(2|q_{13}|^2 -|q_{12}|^2)+4 q_{13}^2 q_{14}^{*}- 4
q_{13}^2 q_{12}^{*}-4 q_{12}^2 q_{14}^{*}=0 \nonumber
\end{eqnarray}
The Hamiltonian of (\ref{eq:D4.5}) is obtained from the general
expression (\ref{eq:MNLSham1}) by imposing  the reduction
constraints:
\begin{eqnarray}\label{eq: H-D6}
H&=&H_{\rm kin}+H_{\rm int}, \nonumber \\ H_{\rm kin}&=&{3\over
a}\int_{-\infty}^{\infty}\rd x\,\left(
 {\partial q_{12} \over
\partial x}{\partial  q_{12}^{*} \over
\partial x} + 2 {\partial  q_{13}  \over \partial
x}{\partial  q_{13}^{*} \over
\partial x}  + 2 {\partial  q_{14}  \over \partial
x}{\partial  q_{14}^{*} \over
\partial x} \right),\\
H_{\rm int}&=&{3\over a}\int_{-\infty}^{\infty}\rd x\, \left( 6
|q_{12}|^4+ 6|q_{13} |^4+  |q_{14} |^4 + 8 |q_{14}|^2( |q_{12}|^2+
|q_{13}|^2)
 \right) \nonumber\\
&+& \frac{6}{a}\int_{-\infty}^{\infty}\rd x\, (q_{12}^2
q_{14}^{*2}+ q_{14}^2 q_{12}^{*2}-q_{13}^2 q_{14}^{*2}- q_{14}^2
q_{13}^{*2}-q_{12}^2 q_{13}^{*2}- q_{13}^2 q_{12}^{*2}),\nonumber
\\
&+& \frac{12}{a}\int_{-\infty}^{\infty}\rd x\, ( (|q_{12}|^2-2
|q_{13}|^2)( q_{12} \,q_{14}^{*}+ q_{14} \,q_{12}^{*})),\nonumber
\\
&+& \frac{12}{a}\int_{-\infty}^{\infty}\rd x\, (q_{12}q_{14}
q_{13}^{*2}+q_{13}^2 q_{14}^{*} q_{12}^{*})),\nonumber
\end{eqnarray}
The soliton solutions of the reduced MNLS require additional
efforts. The problem is that the generic expression for the
dressing factor (\ref{eq:4.2}) does not satisfy the reduction
conditions.

\section{Reductions and the scattering data}\label{sec:Rscd}

The reduction conditions (\ref{eq:3.1}) and (\ref{eq:3.3}) imposed
on the potential of the Lax operator (\ref{eq:Lax-MNLS}) induce an
invariance conditions for the corresponding fundamental analytical
solutions
\begin{eqnarray}
&&\mbox{\bf Type I:}\quad  B^{-1}(\chi^+(x,t,\lambda^*))^\dag
B=\left(\begin{array}{cc} c^- (\lambda)& 0 \\
0 & a^- (\lambda)\end{array}\right)
(\chi^-(x,t,\lambda))^{-1}, \\
&&\mbox{\bf Type II:}\quad
C^{-1}(\chi^\pm(x,t,\lambda^*))^*C=\chi^\pm(x,t,\lambda),
\end{eqnarray}
and for the scattering matrix (\ref{eq:ScM})
\begin{eqnarray}
&&\mbox{\bf Type I:}\quad
B^{-1}T^\dag(t,\lambda^*)B=(T(t,\lambda))^{-1}, \\
&&\mbox{\bf Type II:}\quad  C^{-1}T^*(t,\lambda^*)C=T(t,\lambda).
\end{eqnarray}
If we represent the internal automorphisms $S_\alpha$ of {\bf type
I} in the form
\[
S_\alpha= \left(
\begin{array}{cc}
B_+ & 0\\
0 & B_-\\
\end{array}
\right)
\]
the reduction conditions on the matrix blocks $a^\pm(\lambda)$,
$b^\pm(\lambda)$, $c^\pm(\lambda)$ and $d^{\pm}(\lambda)$ in the
scattering matrix and its inverse (\ref{eq:ScM}) read:
\begin{eqnarray}\label{eq:a-c.I.red}
B_+ (a^+(\lambda^*))^\dag B_+^{-1}=c^+(\lambda), \qquad B_+
(b^+(\lambda^*))^\dag B_-^{-1}=d^-(\lambda),\\
B_- (b^-(\lambda^*))^\dag B_+^{-1}=d^+(\lambda), \qquad B_-
(a^-(\lambda^*))^\dag B_-^{-1}=c^-(\lambda).\nonumber
\end{eqnarray}
For the other reductions of {\bf type II} the internal
automorphisms preserving $J$ up to a sign has off-diagonal block
structure as follows:
\[
S_\alpha= \left(
\begin{array}{cc}
0 & C_+\\
C_- &0\\
\end{array}
\right)
\]
and the matrix blocks of $T(\lambda)$ are constrained by
\begin{eqnarray}\label{eq:a-a.II.red}
C_+ (a^-(\lambda^*))^* C_+^{-1}=a^+(\lambda), \qquad C_-
(a^+(\lambda^*))^* C_-^{-1}=a^-(\lambda),\\
C_+ (b^+(\lambda^*))^* C_-^{-1}=-b^-(\lambda), \qquad C_-
(b^-(\lambda^*))^* C_+^{-1}=-b^+(\lambda).\nonumber
\end{eqnarray}
Similar reduction constraints could be written also for the inverse
$\hat{T}(\lambda)$ to the scattering matrix $T(\lambda)$.

Our last remark here is that the reductions described above can not be
applied to any generic NLEE. Indeed, eq. (\ref{eq:evol}) will be
compatible with the reduction only if the dispersion law $f(\lambda ) $
satisfies
\begin{equation}\label{eq:disp-f}
B_+^{-1}\left(f_{+}(\lambda^{*})\right)^{\dagger}B_+=f_{+}^{-1}(\lambda)
\qquad
B_-^{-1}\left(f_{-}(\lambda^{*})\right)^{\dagger}B_-^{-1}=f_{-}^{-1}(\lambda).
\end{equation}
Any generic NLEE is compatible with reduction of Type II if the
dispersion law complies with
\begin{equation}\label{eq:disp}
C_+\left(f_{-}(\lambda^{*})\right)^{*}C_+^{-1}=f_{+}(\lambda)
\qquad
C_-\left(f_{+}(\lambda^{*})\right)^{*}C_-^{-1}=f_{-}(\lambda).
\end{equation}

\section{Conclusions}\label{sec:6}

We have described new systems of MNLS type obtained as $\bbbz_4$
and $\bbbz_6 $-reductions of the MNLS related to a ${\bf D.III}$
type symmetric space. The Hamiltonian formalism and the theory of
$\Lambda$-operators for MNLS related to the relevant simple Lie
algebras are briefly discussed. We show how the method, presented
in \cite{1} for the $N$-wave equations and their gauge equivalent
systems can be extended to MNLS type systems \cite{vgn2}. The
reduction of the multi-component nonlinear Schr\"{o}dinger (NLS)
equations on symmetric space ${\bf C.I}\simeq {\rm Sp(2p)}/{\rm
U(p)}$ for $p=2 $ is related to spinor model of Bose-Einstein
condensate. Other interesting reductions of MNLS type equations
were reported in \cite{MV} and a systematic study of the problem
is on the way.

These results can be extended and the reductions of MNLS-type
equations related to other symmetric and homogeneous spaces can be
explored. As a result one can systematically obtain and classify
new integrable systems of MNLS type. The method is explicitly
gauge covariant and can also be applied to their gauge equivalent
systems of Heisenberg ferromagnet type. Such research would entail
a voluminous calculations and will be continued in subsequent
publications.

\section*{Acknowledgments}\label{sec:Ack}

This work has been supported also by the National Science
Foundation of Bulgaria, contract No. F-1410.


\begin{thebibliography}{77}

\bibitem{IMW04} Ieda J., Miyakawa T. and Wadati M., {\it Exact
Analysis of Soliton Dynamics in Spinor Bose-Einstein Condensates},
Phys. Rev Lett. {\bf 93}, (2004), 194102.

\bibitem{ForKu*83} Fordy A. P. and Kulish P. P., {\it Nonlinear Schrodinger
Equations and Simple Lie Algebras}, Commun.\ Math.\ Phys.\ {\bf
89} (1983) 427--443.

\bibitem{LLMML05} Li L., Li Z., Malomed B. A., Mihalache D.  and
Liu W. M., {\it Exact Soliton Solutions and Nonlinear Modulation
Instability in Spinor Bose-Einstein Condensates}, Phys. Rev. A
{\bf 72}, (2005) 033611.

\bibitem{Helg} Helgasson S.,
{\it Differential Geometry, Lie Groups and Symmetric Spaces},
Academic Press, (1978).

\bibitem{Loos}  Loos O, {\it Symmetric Spaces. I: General Theory}, W. A. Benjamin,
Inc., New York-Amsterdam, 1969;\\
{\it Symmetric spaces. II: Compact Spaces and Classification}, W.
A. Benjamin, Inc., New York-Amsterdam, 1969.


\bibitem{K} Kaup D. J., {\it Closure of the Squared Zakharov-Shabat
Eigenstates}, J.\ Math.\ Annal. Appl.\ {\bf 54} (1976)
849--864;\\
 Gerdjikov V. S. and Khristov E. Kh., {\it On the Evolution
Equations Solvable with the Inverse Scattering Problem. I. The
Spectral Theory; II. Hamiltonian Structures and B\"acklund
Transformations}, Bulg. J. Phys. {\bf 7} (1980) No.1 28--41; {\bf
7} (1980) No.2 119--133.


\bibitem{G*86} Gerdjikov V. S. and Kulish P. P., {\it The Generating Operator
for the $n\times n$ Linear System},
Physica D {\bf 3} (1981) 549--564. \\
 Gerdjikov V. S., {\it Generalized Fourier Transforms for the
Soliton Equations. Gauge Covariant Formulation}, Inverse Problems
{\bf 2} (1986) 51--74.


\bibitem{VSG*94} Gerdjikov V. S., {\it The Generalized Zakharov--Shabat System
and the Soliton Perturbations}, Teor. Mat. Fiz. {\bf 99} (1994)
292-299.

\bibitem{MV} Gerdjikov V. S., Grahovski G. G. and Kostov N. A.,
{\it On the Multi-component NLS Type Equations on Symmetric
Spaces. Reductions and Scattering Data Properties. } Report at the
International Conference "Contemporary Aspects of Astronomy,
Theoretical and
Gravitational Physics", May, 20 - 22, 2004, Sofia, Bulgaria;\\
Gerdjikov V. S., Grahovski G. G. and Kostov N. A., {\it On the
Multi-component NLS Type Equations on Symmetric Spaces. $\bbbz_2
$-Reductions and Soliton Solutions. } Report at the Sixth
International conference "Geometry, Integrability and
Quantization", July 3--10, 2004, Varna, Bulgaria, Sofia, Softex
(2005) 203--217.


\bibitem{VSG-pres} Gerdjikov V. S., {\it Basic Aspects of Soliton
Theory}, Review article in "Geometry, Integrability and
Quantization - VI", Sofia, Softex (2005) 78--122.


\bibitem{Za*Mi} Zakharov V. E. and Mikhailov A. V., {\it On the
Integrability of Classical Spinor Models in Two--dimensional
Space--time}, Commun. Math. Phys. {\bf 74} (1980) 21--40.

\bibitem{2} Mikhailov A. V., {\it The Reduction Problem and the
Inverse Scattering Problem}, Physica D {\bf 3} (1981) 73--117.

\bibitem{AKNS} Ablowitz M., Kaup D., Newell A. and Segur H., {\it The
Inverse Scattering Transform -- Fourier Analysis for Nonlinear
Problems}, Stud. Appl. Math. {\bf 53} (1974) 249--315.

\bibitem{I04} Ivanov R., {\it On the Dressing Method for the
Generalized Zakharov-Shabat System}, Nucl. Phys. {\bf B 694},
(2004) 509--524.

\bibitem{FaTa} Faddeev L. D. and Takhtadjan L. A., {\it Hamiltonian
Approach in the Theory of Solitons\/}, Springer Verlag, Berlin,
(1987).

\bibitem{ZMNP} Zakharov V. E., Manakov S. V., Novikov S. P.  and
Pitaevskii L. I. , {\it Theory of Solitons. The Inverse Scattering
Method}, Plenum Press (Consultant Bureau), N.Y., (1984).


\bibitem{CaDe} Calogero F. and Degasperis A., {\it Nonlinear Evolution
Equations Solvable by the Inverse Spectral Transform I,II}, Nuovo
Cim.\ {\bf 32B} (1976) 201--242; {\bf 39B} (1976) 1--54.

\bibitem{grah:wildgelf} Gelfand I. M. and Dickey L. A., {\it Asymptotics
Properties of the Resolvent of Sturm--Liouville Equations and the
Algebra of KdV Equations}, Usp. Mat. Nauk {\bf 30} (1977)
67--100;\\
Gelfand I. M. and Dickey L. A., {\it Resolvents and Hamiltonian
Systems}, Funk. Anal. Pril. {\bf 11}, n.2, 11--27 (1976).


\bibitem{LMP} ~Gerdjikov V.~S., {\it On the Spectral Theory of the
Integro--ifferential Operator Generating Nonlinear Evolution
Equations}, Lett. Math. Phys. {\bf 6} (1982) 315--324.

\bibitem{CD2} Calogero F. and Degasperis A., {\it Coupled
Nonlinear Evolution Equations Solvable via the Inverse Spectral
Transform and Solitons that Come Back: the Boomeron}, Lett. Nuovo
Cimento \textbf{16} (1976) 425--433.


\bibitem{D1} Degasperis A. , {\it Solitons, Boomerons,
Trappons}, In: Nonlinear Evolution Equations Solvable by the
Spectral Transform, edited by F. Calogero, Pitman, London, 1978,
pp. 97--126.


\bibitem{CaDe-new} Calogero F. and Degasperis A., {\it New Integrable
Equations of Nonlinear Schr\"{o}dinger Type}, Stud. Appl. Math.
{\bf 113} (2004) 91--137.


\bibitem{1} Gerdjikov V. S., Grahovski G. G., Ivanov R. I. and Kostov N. A.,
{\it $N $-wave Interactions Related to Simple Lie Algebras.
$\bbbz_2$- reductions and Soliton Solutions}, Inv. Problems {\bf
17} (2001) 999--1015;\\
Gerdjikov V. S., Grahovski G. G. and Kostov N. A., {\it Reductions
of $N $-wave Interactions Related to Low--rank Simple Lie
Algebras. I: $\bbbz_2$--reductions}, J. Phys. A: Math. Gen. {\bf
34} (2001) 9425--9461.

\bibitem{DrSok} Drinfel'd, V and Sokolov, V V,
{\it Lie Algebras and Equations of Korteweg - de Vries Type}, Sov.
J. Math. {\bf 30} (1985) 1975--2036.


\bibitem{vgn2} Gerdjikov V. S., Grahovski G. G. and Kostov N. A., {\it On
$N$-wave Type Systems and their Gauge Equivalent}, Eur. Phys. J. B
{\bf 29} (2002) 243--248.


\bibitem{MiLo1} Lombardo S. and Mikhailov A. V., {\it Reductions of
Integrable Equations. Dihedral Group}, J. Phys. A: Math. Gen. {\bf
37} (2004) 7727--7742.

\bibitem{MiLo2} Lombardo S. and Mikhailov A. V., {\it Reduction Groups and
Automorphic Lie Algebras}, Commun. Math Phys. {\bf 258} (2005)
179--202.



\end{thebibliography}
\end{document}